\documentclass[preprint,aps,nofootinbib]{revtex4}
\usepackage{graphicx}
\usepackage{epstopdf}
\usepackage{subfigure}
\usepackage{textcomp}
\def\bkR{{\rm I\kern-.17em R}}
\def\bkC{{\rm \kern.24em \vrule width.05em height1.4ex depth-.05ex \kern-.26em C}}

\def\al{\alpha}
\def\be{\beta}
\def\ga{\gamma}
\def\de{\delta}

\def\th{\theta}

\def\la{\lambda}

\def\si{\sigma}

\def\om{\omega}
\def\Ga{\Gamma}

\def\Om{\Omega}
\def\mn{{\mu\nu}}

\def\frac#1#2{{\textstyle{{#1}\over {#2}}}}

\def\lsim{\mathrel{\rlap{\lower4pt\hbox{\hskip1pt$\sim$}}
    \raise1pt\hbox{$<$}}}
\def\gsim{\mathrel{\rlap{\lower4pt\hbox{\hskip1pt$\sim$}}
    \raise1pt\hbox{$>$}}}
\def\sqr#1#2{{\vcenter{\vbox{\hrule height.#2pt
         \hbox{\vrule width.#2pt height#1pt \kern#1pt
         \vrule width.#2pt}
         \hrule height.#2pt}}}}

\def\laq{\raise 0.4 ex \hbox{$<$}\kern -0.8 em\lower 0.62 ex\hbox{$\sim$}}
\def\gaq{\raise 0.4 ex \hbox{$>$}\kern -0.7 em\lower 0.62 ex\hbox{$\sim$}}

\def\be{\begin{equation}}
\def\ee{\end{equation}}
\def\ba{\begin{eqnarray}}
\def\ea{\end{eqnarray}}

\def\dalemb#1#2{{\vbox{\hrule height.#2pt
        \hbox{\vrule width.#2pt height#1pt \kern#1pt \vrule width.#2pt}
        \hrule height.#2pt}}}

\def\dalemb#1#2{{\vbox{\hrule height.#2pt
        \hbox{\vrule width.#2pt height#1pt \kern#1pt \vrule width.#2pt}
        \hrule height.#2pt}}}

\def\gtorder{\mathrel{\raise.3ex\hbox{$>$}\mkern-14mu
             \lower0.6ex\hbox{$\sim$}}}
\def\ltorder{\mathrel{\raise.3ex\hbox{$<$}\mkern-14mu
             \lower0.6ex\hbox{$\sim$}}}

\begin{document}

\rightline{DF/IST-7.2010}
\rightline{May 2010}

\title{The Spirit of Unification: The Wei of Physics\footnote{Talk delivered at the ``Mira Fernandes and his age: 
An historical Conference in honour of Aureliano de Mira Fernandes (1884-1958)", 17 June 2009, Instituto Superior T\'ecnico, 
Lisbon, Portugal.}

}

\author{Orfeu Bertolami\footnote{Also at Instituto de Plasmas e Fus\~ao Nuclear,
IST. E-mail: orfeu@cosmos.ist.utl.pt}}

\vskip 0.3cm

\affiliation{Departamento de F\'\i sica, Instituto Superior T\'ecnico \\
Avenida Rovisco Pais 1, 1049-001 Lisboa, Portugal
}


\vskip 3cm

\begin{abstract}

\vskip 1cm

{We review, in a historical perspective, developments in physics which led to
the emergence of unifying ideas and theories. Some attention is paid
to the theoretical programme that started in the second decade of the XXth
century and whose objective was to reach a unified description of gravity and
electromagnetism. These attempts can be regarded as conceptually akin to the
contemporary ``theories of everything" which aim to unify all interactions of Nature.
}

\end{abstract}

\maketitle

\section{Introduction}

\vskip 0.3cm

\centerline{{\bf Unifying ideas in science and philosophy: a chronology}}

\vskip 0.35cm

Many philosophical systems aimed at unifying the knowledge of their time 
about the physical world. In physics this trend has been more visible from the 
second decade of the  XXth century onwards, when the first attempts to unify 
gravity and electromagnetism were developed. However, one should realize 
that the search for unifying principles is at the very core of the development 
of physics since early on. 

In what follows we present a brief chronology of some of the most relevant unifying ideas in physics and philosophy:  
 
\noindent
{\bf ca. 440 BC}: The Pre-Socratics Greek philosophers Leucippus and Democritus were the first to put forward the idea 
that {\it everything is composed of atoms} and that {\it atoms are physically indivisible and in constant motion in empty space}. 
This is a quite elegant solution to the problem of maintenance of identity in a world in constant transformation.
\vspace{0.3cm}

\noindent
{\bf 384-322 BC}: Aristotle's Physics, where {\it physis} means Nature, treats the problem of growth 
and change in the whole natural world, which includes plants, animals and objects. Aristotle also unifies motion and movement, 
which manifest themselves through the change in quality and in quantity. 
\vspace{0.3cm}

\noindent
{\bf ca. 630 AD}: Islam arises as a religious movement that claims to unify the Abrahamic monotheistic religions.
\vspace{0.3cm}

\noindent
{\bf ca. 1650}: Emergence of the scientific revolution that laid the basis of Newton's mechanics and the law of universal gravitation, whose
applicability everywhere in the Universe made possible systematizing the knowledge of the whole physical world.
\vspace{0.3cm}

\noindent
{\bf ca.  1850}: French philosopher Auguste Comte put forward his Positivist Philosophy as, in his view, humankind was living in the scientific phase of its 
history and as such all sciences, natural and human, should be regarded as a whole.
\vspace{0.3cm}

\noindent
{\bf ca. 1870}: Karl Marx's historical materialism aimed to understand the whole human history 
through the economical relationships present in society.
\vspace{0.3cm}

\noindent
{\bf 1859}: Charles Darwin (and Alfred Russell Wallace) put forward the theory of evolution and the precept of survival of the 
fittest, which should apply to all living organisms. 
Darwin's theory has influenced many thinkers, and his ideas were often improperly used, most particularly in social sciences. 
A few decades later, Ludwig Boltzmann outlined his idea of {\it theoretical pluralism}, according to which  scientific theories 
should be regarded as representations of Nature that should ``compete for their survival'', like living organisms. 
According to Boltzmann, the ultimate scientific knowledge was unattainable --- a remarkably modern view, quite akin 
to Popper's view that scientific theories are actually no more than conjectures that remain valid as far as they account 
for the observational facts.  
\vspace{0.3cm}

\noindent
{\bf 1865}: Maxwell's equations unify electricity, magnetism, optics and electromagnetism.
\vspace{0.3cm}

\noindent
{\bf 1869}: Mendeleyev's periodic system explains all chemical bounds through the common properties 
of elements (92 natural ones) properly gathered according to the number of electrons available for a chemical reaction.

\vskip 0.4cm

\centerline{{\bf The Relativity Revolution}}

\vskip 0.4cm

\noindent
{\bf 1905}: Einstein's Special Relativity (SR) unifies electromagnetism and a generalized mechanics, an approach 
that allows for the whole physics to be expressed in a frame independent way and invariant under the Lorentz transformations.
\vspace{0.3cm}

\noindent
{\bf 1908}: Minkowski shows that SR is more meaningful with a further unification of space and  time so to treat physical 
phenomena in a space-time continuum (see Refs. \cite{Bertolami2008,BertolamiLobo2009} for thorough discussions).
\vspace{0.3cm}

\noindent
{\bf 1915}: Einstein's General Relativity unifies SR, gravity and Riemannian differential geometry imposing that physics 
should be invariant under 
general coordinate transformations, {\it i.e.}, diffeomorphisms. This is achieved by promoting the space-time metric, $g_{\mu\nu}$, which 
determines distances in space-time  
\be\label{eq1.1}
ds^2=g_{\mu\nu}dx^{\mu}dx^{\nu}~,
\ee
into a dynamical variable that solves the so-called Einstein's 
field equations for a given matter-energy configuration described by the 
energy-momentum tensor $T_{\mu\nu}$:
\be\label{eq1.2}
R_{\mu\nu} - {1 \over 2}g_{\mu\nu}R= {8 \pi G \over c^4}T_{\mu\nu}~,
\ee 
where $R_{\mu\nu}$ is the Ricci tensor, obtained by a contraction of the 
Riemann tensor, the fundamental tensor of differential geometry to detect curvature, 
$R = R^{\lambda}_{\lambda}$, $G$ is Newton's constant and $c$ the speed of 
light. 

General Relativity assumes that space-times admits a symmetric and metric compatible connection, $g_{\mu\nu;\lambda}=0$, the  
Levi-Civita connection. The Riemann tensor is built with this connection and the contracted version of the 
Bianchi identity implies that the energy-momentum tensor is covariantly conserved.   
This formulation emphatically states that physics is a science of space-time (see Ref. \cite{Bertolami2006a} for an extensive 
discussion). Furthermore, the theory 
accounts for all known observational facts, from the solar system to the largest observed scales --- provided 
that in the latter, dark energy and dark matter are included into the cosmological description (see Ref. \cite{Bertolami2006b} 
for a thorough discussion).   

Somewhat later, in 1917, while discussing for the first time the cosmological implications of GR, Einstein \cite{Einstein1917} 
rejected a legitimate solution of his equation that predicted the expansion of the Universe and, rather unfortunately, 
concluded that the field 
equations were incomplete and should be implemented with a constant term, the cosmological constant, $\Lambda$:
\be\label{eq1.3}
R_{\mu\nu} - {1 \over 2}g_{\mu\nu}R + \Lambda g_{\mu\nu}= {8 \pi G \over c^4}T_{\mu\nu}~.
\ee  
One could assume that the observed expansion of the Universe would render the cosmological constant redundant. However, 
the recently detected acceleration on the expansion of the Universe \cite{Perlmutter} brings back the cosmological constant, which 
suitably adjusted, is the simplest possible explanation for this observational fact.

\vskip 0.4cm

\centerline{{\bf Unitary Field Theories\footnote{The word unitary is used in this section for historical reasons. It means unified and not unitary in the mathematical sense.} }}

\vskip 0.4cm

\noindent
{\bf 1918}: Weyl's dilatation unitary field theory (see Refs. \cite{Goenner2004,Goenner2006} for thorough discussions).
\vspace{0.3cm}

\noindent
Weyl admits that the metric is not compatible, which is expressed via a scalar field, $\la(x)$, assuming the following properties
\begin{eqnarray}
g_{\mn;\la} &:= & Q_{\mn \la}\neq 0 \\
\bar{g}_\mn &=& \la(x) g_\mn~~~~,~~~~ \la(x) \in C^\infty (R) \\ 
Q_\mn^\la &=& -g_\mn Q^\la~~~~,~~~~Q=Q_\la dx^\la \\
Q_\la &\rightarrow & \bar{Q}_\la + {\partial ln \la(x) \over \partial x^\la} \\
dQ &=& \left( {\partial Q_\la \over \partial x^\rho} -  {\partial Q_\rho \over \partial x^\la} \right) dx^\rho \wedge d x^\la := F_{\rho\la} dx^\rho \wedge d x^\la \\
V^\mu_{;\nu} &=& V^\mu_{,\nu} + \Ga_{\nu \rho}^\mu V^\rho ~~~~,~~~~~
\om_{\mu;\nu} = \om_{\mu,\nu} + \Ga_{\nu \mu}^{'\rho}\om_\rho \\\Ga_{\nu \rho}^{'\la} &=& -\Ga_{\nu \rho}^\la
\end{eqnarray}
and then imposes the identification with the electromagnetic field $Q_\rho \leftrightarrow A_\rho$.

\noindent 
His generalized Riemann curvature is built from his new connection 
\begin{eqnarray}
K^\la_{\rho[\mn]} := \partial_\mu \Ga_{\nu \rho}^\la + \Ga_{\mu \th}^\la \Ga_{\nu \rho}^\th - \partial_\nu \Ga_{\mu \rho}^\la - \Ga_{\nu \th}^\la \Ga_{\mu \rho}^\th~,
\end{eqnarray}
yielding the Ricci tensor
\begin{eqnarray}
K^\la_{\mu\nu\lambda} := \partial_\mu \Ga_{\nu \la}^\la + \Ga_{\mu \th}^\la \Ga_{\nu \la}^\th - \partial_\nu \Ga_{\mu \la}^\la - \Ga_{\nu \th}^\la \Ga_{\mu \la}^\th
\end{eqnarray}
and the scalar curvature:
\begin{eqnarray}
K(g,\Ga)=g^\mn K_\mn ~~~~,~~~~K_\mn := K_{\mn \la}^\la~.
\end{eqnarray}

\noindent 
One could also consider theories with a non-symmetric connection, so that {\it torsion} is non-vanishing: 
\begin{eqnarray}
\label{defS}
S_\mn^\la := \Ga_{[\mn]}^\la ~~~~,~~~~ S_\mu = S_{\mu \rho}^\rho ~,
\end{eqnarray}
and the following identity among the relevant quantities is satisfied:
\begin{eqnarray}
V_\mn + K_{[\mn]} = -4S_{[\mu;\nu]} + 2 S_{\mn;\la}^\la + 4S_\mn^\la S_\la ~.
\end{eqnarray}
Clearly, if the connection is symmetric
\begin{eqnarray}
\Ga_{\mn}^\la = \Ga_{(\mn)}^\la \leftrightarrow S_\mn^\la = V_\mn = K_{[\mn]} = 0 ~,
\end{eqnarray}
and one recovers the Levi-Civitta connection which can be expressed in terms of the Christoffel symbols:
\begin{eqnarray}
\Ga_\mn^\la = \left\{ ^{~\la}_\mn \right\}:= {1 \over 2} g^{\la \si}(h_{\mu \si , \nu} + h_{\nu \si , \mu} - h_{\mu \nu , \si}) ~,
\end{eqnarray}
where one has written the most general metric-like tensor as 
$g_\mn = h_\mn + k_\mn$, where $h_\mn:=g_{(\mn)}:=\frac{1}{2} (g_\mn + g_{\nu\mu})$ and $k_\mn:=g_{[\mn]} :=\frac{1}{2} (g_\mn - g_{\nu\mu}) $. 

\vspace{0.3cm}

\noindent
{\bf 1919, 1925}: Kaluza-Klein approach 

\noindent
The Finish physicist Gunnar Nordstr\"om was the first to speculate, in 1909, that space-time could have more 
than four dimensions. A concrete construction based on this idea was put forward by Theodor Kaluza in 1919, who showed that an unified 
theory of gravity and electromagnetism could be achieved through a 5-dimensional version 
of general relativity, provided the extra dimension was compact and fairly small, and could therefore have passed undetected. 
This approach was reexamined by Oskar Klein in 1925, who showed that after correcting some of Kaluza's assumptions, 
the resulting 4-dimensional theory contained general relativity, electromagnetism and a scalar field theory.

This approach was very dear to Einstein till late in his life \cite{Einstein1952}, and has been followed in the 
most important attempts to unify all known four interactions of Nature. In a more modern language, the basic underlying feature of the Kaluza-Klein approach is the assumption 
that the Universe is described by Einstein's general relativity in a $5$-dimensional Riemannian manifold, $M_5$, that is  a product of a 
$4$-dimensional Riemannian manifold, $M_4$, our world, and tiny one-dimensional sphere, $S_1$.

Much later, in the 1970's, it was shown the impossibility of encompassing larger gauge groups than $U(1)$ in $4$ 
dimensions starting from $D$-dimensional general relativity. Furthermore, besides the impossibility of this ``monistic'' approach, {\it i.e.}, 
the route from higher-dimensional gravity down to $4$-dimensional gravity plus gauge theories, some ``no-go theorems'' have 
shown that no non-trivial gauge field configurations could be obtained in $4$ dimensions from D-dimensional gravity theory 
without higher-curvature terms (see {\it e.g.} Ref. \cite{FGW} for the case of $N=1$ supergravity in $D=d+4=10$ dimensions).    
Another quite relevant result is that, in order to obtain chiral fermions in our $4$-dimensional world, $D$ must be even if all 
extra dimensions are compact \cite{Witten1}. 

Actually, in order to get consistent effective $4$-dimensional models arising for instance from a $D$-dimensional 
Einstein-Yang-Mills theory, multidimensional universes of the form $M_D = {\bf R} \times G^{{\rm ext}}/H^{{\rm ext}} 
\times G^{{\rm int}}/H^{{\rm int}}$, should be considered, where $G^{\rm ext(int)}$ and $H^{{\rm ext(int)}}$ are
respectively the isometry groups in 3($d$) dimensions. This technique, known as {\it coset space dimensional reduction} \cite{HM} 
(see Ref. \cite{KMRV} for an extensive discussion), is quite powerful and has many applications in theoretical physics. 
For example, in cosmology, when considering homogeneous and isotropic models (a $1$-dimensional problem), it allows 
for obtaining consistent effective models 
arising from $4$-dimensional \cite{BMPV} or $D$-dimensional Einstein-Yang-Mills-Higgs theories 
\cite{BKM}. In this particular instance, one considers $G^{\rm ext(int)} = SO(4)~(SO(d+1))$ and $H^{{\rm ext(int)}} = SO(3)~(SO(d))$ as
the homogeneity and isotropy isometry groups in $3$($d$) dimensions.

We shall further discuss the more recent efforts based on the Kaluza-Klein approach below.  
Let us now get back to the historical discussion of the unitary field theories. 

\vspace{0.3cm}

\noindent
{\bf 1921}: Eddington's Affine Geometry approach assumes a symmetric connection:
\begin{eqnarray}
\Ga_\mn^\la = \Ga^\la_{(\mn)}
\end{eqnarray}
so that for a constant, $\lambda$, 
\begin{eqnarray}
g_\mn = \la K_{(\mn)}~~~~,~~~~K_{(\mn)}=K^\la_{\mn \la}
\end{eqnarray}
together with a metric compatibility:
\begin{eqnarray}
\nabla_\mu K_{(\nu \la)}(\Ga) = 0 ~.
\end{eqnarray}

Electromagnetism, that is, Maxwell's non-homogeneous Eqs., $F^{\mu\nu}_{;\nu} = J^\mu$, arise from the identification of the 
gauge field strength:
\begin{eqnarray}
F^\mn & := & g^{\mu \la} (\Ga) g^{\nu \rho} (\Ga) K_{[\la\rho]} (\Ga) \\ 
F_\mn &=& \partial_\mu \Ga_\nu - \partial_\nu \Ga_\mu ~,
\end{eqnarray}
with a generalized Ricci tensor:
\begin{eqnarray}
K_\mn (\Ga) = R_\mn (\Ga) + F_\mn (\Ga) ~.
\end{eqnarray}

\noindent
The conservation of charge $J^\mu_{;\mu}=0$ is achieved through the condition (cf. Eq. (\ref{defS})): 
\begin{eqnarray}
S_{\mu \rho}^\la = 0 ~.
\end{eqnarray}

In 1923, Einstein proposed an action for Eddington's approach:
\begin{eqnarray}
{\cal L} = 2 \sqrt{-det(K_\mn)} ~,
\end{eqnarray}
with 
\begin{eqnarray}
\la^2 K_\mn = g_\mn + \phi_\mn
\end{eqnarray}
where $g_{\mu \nu}=g_{(\mu \nu)}$, $\phi_{\mu \nu}=\phi_{[\mu \nu]}$ and  
\begin{eqnarray}
 \phi_\mn:={1 \over 2} \left( {\partial \Ga_{\mu \la}^\la \over \partial x^\nu} - {\partial \Ga_{\nu \la}^\la \over \partial x^\mu} \right) ~.
\end{eqnarray}
It is rather interesting that the relevant quantities are obtained through a variation with respect 
to the contracted connection, $\Ga_{\mu \la}^{\la}$. Approaches where metric and connection are regarded independently in the variational problem are usually 
referred to as {\it first order formalism}. Einstein assumes that the electromagnetic field strength and current are given respectively by:  
\begin{eqnarray}
f^\mn  := {\de {\cal L} \over \de \phi_\mn} ~~~~,~~~~j^\mu := {\partial \hat{f}^\mn \over \partial x^\nu} ~.
\end{eqnarray}
In the same year, there followed another proposal by Einstein in order to get a more encompassing approach, 
which reflects somehow a sign of his dissatisfaction with the whole procedure \cite{Goenner2004,Goenner2006}.
\vspace{0.3cm}

\noindent
{\bf 1922-1923, 1928}: Independently, Cartan and Einstein developed the distant parallelism approach, in  
which torsion plays a central role. One introduces the tetrad or {\it vierbein} field, $e^{\mu}_{\al} $, 
that relates the flat tangent space to the 
Riemannian space metric through the relationship:  
\begin{eqnarray}
g^{\mu \nu} = e^{\mu}_{\al}  e^{\nu}_{\beta} \eta^{\al \beta}  ~,
\end{eqnarray}
so that 
\begin{eqnarray}
e^{\la}_{\al;\mu}=0  ~.
\end{eqnarray}
The connection is given by
\begin{eqnarray}
\Ga_\mn^\la = e^\la_\al {\partial e^\al_\nu \over \partial x^\nu} ~,
\end{eqnarray}
and torsion by the antisymmetric combination 
\begin{eqnarray}
S_\mn^\al=e^\la_\al \partial_{[\mu}{e^\al}_{\nu]} ~.
\end{eqnarray}
The proposed Lagrangian density is the following
\begin{eqnarray}
{\cal L} =\varrho g^\mn S_{\mu \rho}^\la S_{\nu\la}^\rho ~~~~,~~~~ \varrho=det(e^\la_\al)=\sqrt{-det(g_\mn)} ~,
\end{eqnarray}
whose physical meaning is elusive.

Despite their limited success, these ideas allowed Cartan to obtain a set of equations, today referred to as Maurer-Cartan structural 
equations, which are nowadays used as the standard technique to obtain from the basis of 1-forms, $[\om^\nu]$, dual to the tetrad, the torsion 1-form, $S^\mu$, and curvature 2-form, $R^\mu_\nu$, 
for a generic connection that has symmetric as well as antisymmetric components:
\begin{eqnarray}
&&d\om^\mu + \th^\mu_\nu \wedge \om^\nu = S^\mu~~~~,~~~~ S^\mu := {1 \over 2} T^\mu_{\nu \la} \om^\nu \wedge \om^\la  \\ 
&& R^\mu_\nu = d\th^\mu_\nu + \th^\mu_\la \wedge \th^\la_\nu~~~~,~~~~R_\nu^\mu := {1 \over 2} R^\mu_{\nu\la\rho} \om^\la \wedge \om^\rho ~,
\end{eqnarray}
where one has introduced the connection 1-form,  $\th^\mu_\nu$: 
\begin{eqnarray}
\th^\mu_\nu := \Ga^\mu_{\la \nu} \om^\la ~, 
\end{eqnarray}
and used the well known antisymmetric wedge product. 
\vspace{0.3cm}

\noindent
{\bf 1931-1932}: The Italian mathematician Paolo Straneo from Genova considered 
a generalized connection, so that different contractions of the resulting Riemann curvature would concern 
gravity and electromagnetism, respectively. His set up is the following: 
\begin{eqnarray}
\Ga^{'\la}_\mn = -\Ga_\mn^\la ~~~~,~~~~\Ga_\mn^\la = \left\{ ^{~\la}_\mn \right\} + \Om_\mn^\la ~,
\end{eqnarray}
which yield the following field equations:
\begin{eqnarray}
K_\mn = R_\mn - {1 \over 2} g_\mn R + 2\left( {\partial \psi_\mu \over \partial x^\nu } - {\partial \psi_\nu \over \partial x^\mu} \right) = T_\mn ~,
\end{eqnarray}
where $\psi_{\mu}$ is the electromagnetic gauge field. The Riemann tensor is found to be
\begin{eqnarray}
R^\la_{\mn \rho} + \Om^\la_{\mu \rho;\nu} + \Om_{\mu \rho}^\al \Om_{\al \nu}^\la = 0
\end{eqnarray}
leading, throuh contraction,  to gravity
\begin{eqnarray}
R_\mn + \Om_{\rho \mu}^\la \Om_{\la \nu}^\rho = 0
\end{eqnarray}
and electromagnetism
\begin{eqnarray}
\Om^\la_{\mu\rho;\la}=0 ~.
\end{eqnarray}
Hence, the issue is the choice of the connection generalization that allows for the desired result.

\vskip 0.4cm

\centerline{{\bf The contribution of Mira Fernandes to the Unitary Field Theory Discussion}}

\vskip 0.4cm

Let us now focus on the contribution of Aureliano Mira Fernandes of Instituto Superior T\'ecnico, in Lisbon. His proposal
appeared in two notes on the Italian journal {\it Rendiconti della Real Accademia dei Lincei}, in 1932 \cite{MF1} and 1933 \cite{MF2} 
and relies on the work of the Italian mathematician Straneo, as discussed above. He considered three possibilities. 
In his first approach he introduced two new fields:
\begin{eqnarray}
&&C_\mn^\la = \Ga_\mn^\la + \Ga^{'\la}_\mn = C_\mu A_\nu^\la \\ &&
\end{eqnarray}
so that 
\begin{eqnarray}
\Ga_\mn^\la = \left\{ ^{~\la}_\mn \right\} + C_\nu A^\la_\mu~~~~,~~~~ \Ga_\mn^{'\la} = - \left\{ ^{~\la}_\mn \right\} ~,
\end{eqnarray}
which lead to Straneo's field equations through the identification with the electromagnetic field, $-2\psi_\mu = C_\mu$, assuming that
\begin{eqnarray}
S_\mn^\la :={ 1 \over 2 } \left( \Ga^\la_\mn - \Ga^{\la}_{\nu\mu} \right)=0~~~~,~~~~S_\mn^{'\la} := { 1 \over 2 } \left( \Ga^{'\la}_\mn - \Ga^{'\la}_{\nu\mu} \right) = 0 ~.
\end{eqnarray}
The above assumption is also considered in his second approach, where $C^\la_\mn = C_\mu A^\la_\nu$, and:
\begin{eqnarray}
&&\Ga_\mn^\la = \left\{ ^{~\la}_\mn \right\} ~~~~,~~~~\Ga_\mn^{'\la} = -\left\{ ^{~\la}_\mn \right\} + C_\nu A^\la_\mu ~,
\end{eqnarray}
which allows for obtaining Straneo's field equations through the identification with the electromagnetic field, $2 \psi_\mu = C_\mu$.

Finally, a third choice involves the combination of the two previous ones:
\begin{eqnarray}
&& \Ga_\mn^\la = \left\{ ^{~\la}_\mn \right\} + \Om_\mn^\la ~~~~,~~~~ \Ga_\mn^{'\la} = -\left\{ ^{~\la}_\mn \right\} + \Om_\mn^{'\la} 
\end{eqnarray}
so that 
\begin{eqnarray}
C_\mn^\la = \Om_\mn^\la + \Om^{'\la}_\mn  ~,
\end{eqnarray}
which allows for satisfying Straneo's field equations if each of the curvatures resulting from the above connections are taken to vanish separately. 

At this point it is interesting to mention Aureliano Mira Fernandes' thoughts on the unification principle. These 
appeared in a book published in 1933, based on lectures he presented 
at the Instituto de Altos Estudos of the Academia das Ci\^encias de Lisboa in February 1st and 4th, 1933, ``Modernas 
Concep\c c\~oes da Mec\^anica"  \cite{MF3}:

\noindent
``One has then reached Riemann's vision according to which physical phenomena should be 
governed by an underlying 
geometrical concept; and conversely, that geometry is determined by the content of space. The unity principle, as a 
scientific and philosophical 
goal, is philosophically boosted and dignified. And in the process, domains of mathematical thinking get better equipped for 
future requirements of the physical theories. ... This is a singular fate for a science whose evolution becomes, 
by rule, more interesting as more detached it gets from the realm of the direct applications!''\footnote{``Atingiu-se o ideal de Riemann da 
subordina\c c\~ao de todos os fen\'omenos f\'isicos a uma concep\c c\~ao geom\'etrica; 
e inversamente, da determina\c c\~ao da geometria pelo conte\'udo espacial.
O princ\'ipio da unidade, como ideal cient\'ifico e filos\'ofico fica filosoficamente robustecido e dignificado. E ficam 
tamb\'em, agora como sempre, os dom\'inios do pensamento matem\'atico sobejamente 
apetrechados para uma poss\'ivel interpreta\c c\~ao de futuras exig\^encias das teorias f\'isicas. ... 
Singular destino este, o duma ci\^encia cuja evolu\c c\~ao \'e, em regra, tanto mais interessante quanto mais 
desinteressada!''.}

Let us take on from this quote and reflect a bit upon the unification principle and add up some of our own thoughts on the matter. 
Unification seems to be a natural trend in science as, from time to time, knowledge 
acquired about a class of natural phenomena allows for unifying theories and models that explain apparently distinct 
phenomena. Unification is achieved in the sense that qualitatively 
distinct phenomenological manifestations can be explained by a common underlying theory or model. This qualitative change is 
possible if, and only if, broad and detailed experimental data are available. Furthermore, the  
unification of theories and models leads to conceptual simplicity, even though most often at expense of mathematical complexity. 

It is important to point out that unification is more than a methodological procedure, as the unity of Nature imposes that 
all its phenomenological manifestations should be regarded as a whole, without artificial divisions, and hence have a 
correspondence with particular elements of any 
serious theoretical framework that aims at describing the most salient features of the Universe.  
From this perspective, and given the lack of decisive phenomenological facts to guide their efforts, it is not surprising that the enterprise 
carried by Einstein, Weyl, Eddington, Cartan and followers  could not be successful in its aim to unify gravity and 
electromagnetism. One could add that their disregard of Quantum Mechanics and its methods to address the nuclear and subnuclear 
interactions was also a highly questionable methodological choice.

Despite these shortcomings, on a broad sense, 
their contributions were nevertheless valuable --- as they threaded their way 
through unknown theoretical landscapes, mapped the encountered fruitless regions and, what is perhaps more relevant, 
devised methods that have been 
useful in many areas of physics. Their failure also illustrates that a purely theoretical  
approach is unlikely to achieve the unification goal through the labyrinth of theoretical and mathematically valid possibilities. 
Phenomenological guidelines are essential and cannot be disregarded in the process of building 
new theories and models. For sure, phenomenology on its own is no more than an inert compilation of data, notwithstanding the recent 
emergence of increasingly powerful computers and huge databases, which have led to the idea that 
models and theories can be built algorithmically ---
most particularly in information and social sciences, and in engineering and management \cite{DC}. 
This author is skeptical whether these methods can be straightforwardly applied to physics, given that, most  
often, physical theories go much beyond the phenomenology that is available at the time they are proposed. It is precisely for these  
reasons that we believe that the spirit of unification works as a sort of taoist {\it Wei} principle, that is to say, an intuition, 
a feeling about ``acting or not acting", about ``doing, but not overdoing" in the process of building new unified physical theories.      

\vskip 0.4cm

\centerline{{\bf The Quantum Revolution}}

\vskip 0.4cm

Let us now turn to another fundamental development in XX's century physics. Somewhat after the development of General Relativity, 
the understanding of 
microphysics (atoms, molecules, nuclear physics and beyond) made possible the emergence of Quantum Mechanics (QM). 
The theory has gone itself through 
a process of methodological unification from 1924 till the 1930's. Indeed, the set of principles that were put forward by 
Bohr and Heisenberg (The Matter-Wave Complementarity Principle and The Uncertainty Principle) have provided abundant insight about 
the need to pursue a fresh new vision of Nature in what concerned microphysical phenomena. Two distinct formulations were then put 
forward: matrix mechanics, proposed by Heisenberg, Born and Jordan in 1925; and, in 1926, 
wave mechanics, based on Schr\"odinger's wave equation. These two formulations were shown to be unitarily equivalent, a result known 
as the Stone-von Neumann theorem. 

In 1928, Dirac presented the relativistic version of the wave equation, and showed that its solutions admit 
particles and {\it antiparticles}, states with the same mass, but with opposite charge and baryon (lepton) number. 
In 1932, the discovery of the positron (the anti-electron) by Carl Anderson did confirm 
the physical validity of Dirac's equation, which unifies QM with SR and the concept of spin. 

Actually, somewhat earlier, in 1926, Dirac suggested that analytical classical mechanics was connected with QM through a formal 
relationship between Poisson's brackets and the commutator of physical observables:
\begin{eqnarray}
\left\{ ~~,~~\right\}_{\rm PB} \rightarrow {1 \over i \hbar} \left[ ~~,~~\right] ~.
\end{eqnarray}
This connection requires the introduction of 
Planck's constant, a distinct feature of QM, and the imaginary unit, which arises as the fundamental 
entity of QM: the wavefunction that solves Schr\"odinger's wave equation is not an observable, being hence undetermined 
up to an imaginary phase (see {\it e.g.} Ref. \cite {Dirac} for an extremely elegant presentation). 

In 1932, Wigner introduced the so-called quasi-probability distribution, while studying the quantum corrections to classical 
statistical mechanics. This distribution allows to link Schr\"odinger's wave function to a probability distribution in phase space, 
as it maps real phase-space functions to Hermitian operators. Thus, a formal unification of 
QM with statistical mechanics is achieved. Later, in 1949, Moyal did  
show that this distribution is the basis of the encoding of all quantum expectation values, and hence quantum mechanics  
in phase space. These developments are the basis of the most recent phase-space noncommutative generalizations 
of QM \cite{Bertolami2005,Bertolami2006c,Bastos2008a}. 

Finally, in 1948, Feynman has developed his path integral formalism and shown its equivalence with QM. 

Other relevant developments towards an unified view of Nature include:

\noindent
{\bf 1934}: Fermi's theory of the weak force describes the interactions through the contact of current densities 
composed by two fermions each, $J_\mu = \bar{\psi} \ga_\mu \psi$, with strength $G_F \simeq 10^{-5}~m_p^{-2}$, where $m_p$ is the 
proton mass and $\ga_\mu $ denotes the Dirac's matrices:
\begin{eqnarray}
&& {\cal L}_W = G_F J_\mu J^\mu ~.
\end{eqnarray}
Fermi's theory is an important precursor of the field theory models that would eventually describe the fundamental interactions.  

\vspace{0.3cm}

\noindent     
{\bf Quantum Electrodynamics (QED)}: In late 1940's and early 1950's, Feynman, Schwinger, Tomonaga, Dyson and others have developed 
the basic tools of quantum field theory and have shown that QED, the quantum field theory of charges and photons, 
was a fully consistent theory that could achieve, 
through the renormalization procedure, an impressive power of predictability (see {\it e.g.} Ref. \cite{Feynman} 
for an insightful account). 
\vspace{0.3cm}

\noindent
{\bf Yang-Mills Theory}: In 1954, Yang and Mills (and independently, Salam and Shaw) did propose a generalization of  
Maxwell's electromagnetism, based on the $U_{EM}(1)$ gauge group, that was invariant under a $G = SU(2)$ non-abelian gauge 
group transformation. 
The resulting Yang-Mills equations are similar to Maxwell's ones and are 
expressed in terms of the field strength, $F^a_{\mu \nu}$, of the gauge field,  $A^a_{\mu}$:
\begin{eqnarray}
&&\partial^\mu F^a_\mn  =  J^a_\nu~~~~,~~~~F^a_{[\mn,\la]}=0~, \\ F^a_\mn & = & \partial_\mu A^a_\nu - \partial_\nu A^a_\mu + g f^a_{bc} A_\mu^b A_\nu^c  
\end{eqnarray}
where $a$ is an internal space index that corresponds to the generators of the gauge group (for the $SU(2)$, $a=1,2,3$), 
$f^a_{bc}$ are the structure constants of the gauge group, $g$ is the coupling constant, and $J^a_\nu$ are the 
current densities of matter, as suggested by Fermi. 

Let us close this subsection by stepping out of physics for a while.

\vspace{0.3cm}

\noindent
{\bf 1956}: Crick and Watson's discovery of the DNA molecular structure made possible the understanding of the hereditary 
mechanism and of the basic features of the protein synthesis of all living organisms. This revolution has led to a 
remarkable progress in biology, the most recent being the Human Genome Project --- concluded in 2003, and carried out by teams 
led by Craig Venter and James Watson. The subsequent sequencing of the genome of other organisms did open the 
perspective of yet another revolution, through the detailed understanding of the protein synthesis and 
the possibility of unraveling an unifying paradigm. Actually, the emergence of a unified biology has been discussed for 
quite some time \cite{Smocovitis}.

\vspace{0.3cm}

\noindent
{\bf 1960's}: Plate tectonics theory enabled, through the work of Wegener, Holmes, Hess, Deitz and others, to understand 
geological phenomena on Earth (and on any rocky planet!) in an encompassing way \cite{Klous}.

\section{Forces of the World, Unite!}

\vskip 0.4cm

\centerline{{\bf The Gauge Principle}}

\vskip 0.4cm

\noindent
{\bf Electroweak (EW) unification}: The unification of the electromagnetic and weak interactions was achieved through the work of 
Glashow (1961), Salam (1968), Weinberg (1967) and others. The main ingredients include the 
gauge group $G = SU_{L}(2) \otimes U_{Y}(1)$ that breaks down, through the Higgs mechanism, to Maxwell's theory $U_{EM}(1)$. The 
fundamental {\it spontaneous symmetry breaking mechanism} that endows the Higgs field with a non-vanishing vacuum expectation value 
(experimentally, $<0|H|0>=246~GeV$) is on its own a 
very interesting example of the unity of Nature, 
as it is the very process underlying phase transitions in condensed matter physics, the Higgs field vacuum 
expectation value playing the role of an order parameter, the Higgs field effective potential being equivalent to Helmholtz's free energy.   
\vspace{0.3cm}

\noindent
{\bf Quantum Chromodynamics (QCD)}: In the 1960-1970's, it has been shown that strong interactions could also be described 
by a gauge theory, QCD, 
with gauge group $SU_c(3)$, where $c$ stands for the strong charge dubbed ``colour". This was achieved through 
the understanding that hadrons could be sorted into groups having similar properties and masses: the ``eightfold way", 
put forward by Gell-Mann and Ne'eman in 1961. Shortly after, Gell-Mann and Zweig proposed that the strong interaction 
group structure should be understood through the existence of three distinct ``colours" of smaller particles inside the hadrons, the 
{\it quarks}. Thus, the basic components of hadrons and mesons are ``coloured" quarks and {\it gluons}, 
the vector bosons of strong interactions, with dynamics ruled by  QCD. The theory has remarkable properties such as 
{\it confinement}, which means that coloured states cannot be directly observed, and 
{\it asymptotic freedom}, 
discovered by David Gross, David Politzer and Frank Wilczek in 1973, which allows for predictions 
of many high energy experiments using the perturbation techniques of quantum field theory. Confinement and asymptotic freedom mean that 
quarks are strongly coupled at low energies and that their coupling gets logarithmically weaker with the growth of energy, respectively.  
 
The description of the electroweak and the strong interactions through the gauge principle gave rise to the so-called 
{\it Standard Model} (SM) of particle interactions, which is so far consistent with most of the particle physics phenomenology. 
However, there exists evidence that physics beyond the SM is needed to account for the 
fact that neutrinos seem to be massive and and that phenomenology requires a new ``sterile'' neutrino 
(see {\it e.g.} Ref. \cite{Winter} for an updated discussion). 
Apart from that, the Higgs boson is the only SM state yet to be detected. The search for the 
Higgs boson field is the main priority of the Large Hadron Collider (LHC).

\vskip 0.4cm

\centerline{{\bf Grand Unified Theories}}

\vskip 0.4cm

The gauge principle and the SM suggest a natural procedure to build more encompassing models. One considers non-abelian gauge theories with symmetry groups that admit 
the SM gauge group as a subgroup. These Grand Unified Theories (GUTs) have been extensively discussed from 1973 onwards by Pati, Salam, 
Georgi, Glashow, Quinn, Weinberg and others (see Ref. \cite{GGRoss} for a thorough discussion), and the most studied cases admitted the following GUT gauge groups and symmetry  
breaking pattern: 
\begin{equation}
GUT := SU(5), SO(10), E(6), ... \rightarrow G_{SM} = SU_c(3) \otimes SU_L(2) \otimes U_Y(1)\rightarrow U_{EM}(1) ~.
\end{equation}
It is important to realize that GUTs are also suggested by the evolution of the coupling constants with energy 
according to the {\it renormalization group equations}. Through these equations it is possible to show that if there is no intermediate 
physics between the EW unification and the GUT, the ``great desert'' hypothesis, for the $SU(5)$ GUT 
\cite{GG} one finds that the unification scale can be as large as, $E_{GUT} \simeq 10^{17}~GeV$ \cite{GQW}. 
Subsequently, precision data arising from LEP collider, the  
$Z$ ``factory" that ran at CERN from 1989 to 2000, have revealed that the coupling 
constants of the electromagnetic, strong and weak interactions would 
meet at about $10^{16}~GeV$, and hence be consistent with a putative unification, 
only in the context of the so-called minimal supersymmetric (see below) extensions of the SM \cite{Amaldi}. This is a further evidence of physics beyond the SM. 

A distinct feature of GUTs is the existence of leptoquark particles whose interactions can mediate violation of baryon and 
lepton numbers. Violation of baryon number, violation of $C$ and $CP$ discrete symmetries, and 
out-of-equilibrium processes, which is a natural feature in an expanding Universe, do allow for the creation of the baryon asymmetry 
of the Universe (BAU). The generation of this asymmetry, usually referred to as {\it baryogenesis}, is vital 
to ensure that the Universe does not end up being composed only by photons, as 
argued by Sakharov in 1967 (see Ref. \cite{Dolgov} for a review). An alternative scenario to achieve the BAU involves 
violation of the baryon number and of the CPT symmetry \cite{Bertolami1997}, which might occur in the context of string theory (see below).

\vskip 0.4cm

\centerline{{\bf The cosmological constant problem}}

\vskip 0.4cm

As discussed above, the spontaneous symmetry mechanism is an essential ingredient of the electroweak 
unification. However, in the process of the Higgs field acquiring a non-vanishing expectation value, the vacuum energy 
becomes non-vanishing and a cosmological constant proportional to $\langle0|H|0\rangle^4$ is generated. 
In quantum field theory, this energy can be disregarded, as only energy differences matter; however, as first pointed 
out by Zel'dovich in 1968 \cite{Zeldovich}, in a realistic setup the vacuum energy gravitates and cannot be 
neglected, as it curves space-time (see also Ref. \cite{Bertolami2009} for an updated discussion). The problem is that, even in the 
most conservative scenario, the SM without any consideration about GUTs, the generated cosmological constant is about $10^{56}$ orders of 
magnitude greater that the value inferred from cosmology, namely $10^{-12}~eV^4$, as pointed out by Linde \cite{Linde}, 
Dreitlein \cite{Dreitlen} and Veltman \cite{Linde} in 1974-1975. The discrepancy is of $O(10^{108})$ if one considers 
GUTs, and as huge as $10^{120}$ in quantum gravity approaches, assuming in this context that the contribution to the 
vacuum energy density is $O(M_P^4)$, where $M_P= \sqrt{\hbar c /G} = 1.2 \times 10^{19}~GeV$ is the Planck mass, 
the typical scale of quantum gravity.

Thus, in order to make sense of the current description of the Universe, 
a cancellation of these vacuum 
contributions must be carried out by terms introduced by hand in the geometrical side of Einstein's field 
equations, Eq. (\ref{eq1.3}), with the corresponding number of decimal places. 
Many solutions for this absurd adjustment problem have been proposed (see Refs. \cite{Bertolami2009,Weinberg,Carroll,Bauer} 
for discussions): for instance, a suppression 
of $10^{-120}$ can be achieved if the vacuum energy density evolves with cosmic time, $t$, as $\rho_V \sim t^{-2}$ \cite{Bertolami1986}.  

The cosmological constant problem is a major gap in our understanding of the unity of Nature, as it indicates that 
the quantum field description of the microscopical world does not match the general relativistic description of 
the Universe. It is a  riddle that 
any fundamental theory must address. Unfortunately, string theory, the most studied quantum gravity approach, has  
not provided a decisive insight on the nature of a possible solution to this difficulty \cite{Witten}. Most recent claims 
about a solution based on anthropic considerations in the context of the {\it landscape} approach (see below) are not consensual.   

\vspace{0.3cm}

\noindent
{\bf Supersymmetry:} After the pioneering work of Wess, Zumino, Deser, van Nieuwenhuizen, Freedman and others in the mid 1970's, 
a fundamental new symmetry has emerged from the drawing board of the theoreticians, {\it supersymmetry}. 
This symmetry relates bosons and fermions with equal masses and prevents 
that radiative corrections in the SM endow the Higgs field with a mass as high as the Planck mass. 
This suggests that new physics beyond the SM is needed in order to bridge the gap between the 
electroweak scale, $M_{EW}\simeq 10^2~GeV$, and the quantum gravity scale. 
This is achieved as supersymmetry allows for a cancellation of the divergencies arising from bosons loops with the 
ones arising from fermions loops. If supersymmetry was an unbroken symmetry, this feature would explain the vanishing 
of the cosmological constant. This is clearly not the case, as known bosons and fermions do not have the same mass. 
This means that the spectrum of elementary particles must be much larger than the one that has been unraveled so far. 

A striking feature of supersymmetry is that the algebra, {\it i.e.}, the anti-commutator, of its generators 
is proportional to the energy-momentum operator. Thus, a local version of supersymmetry corresponds to a general 
coordinate transformation and hence, gravity can be accommodated in a local supersymmetric theory. 
For this very reason, local supersymmetric theories are called supergravity theories 
(see Refs. \cite{Neuiw,Nilles} for extensive reviews). Supergravity is 
an important step towards the unification of gravity with gauge theories. A distinct property of 
supergravity is the presence of one (or more generally, $N \le 8$) partner(s) of the graviton, 
the gravitino, a spin $3/2$ particle that, through the super Higgs mechanism acquires, (in $N=1$ supergravity) a mass \cite{DeserZumino} given by
\begin{equation}
m_{3/2} = \sqrt{{8 \pi \over 3}} {M_{SB}^2 \over M_P}  ~,
\end{equation}
where $M_{SB}$ is the supersymmetry breaking scale. Thus, if supersymmetry is broken at an intermediate scale 
between the electroweak scale and the typical GUT scale, $M_{SB} \simeq 10^{11}~GeV$, one should expect a 
signature of supergravity at the 
LHC collider --- a quite exciting possibility. 

Supersymmetry is also a fundamental ingredient of superstring theory, the most developed approach to understand 
in an unified fashion all interactions of Nature and to harmonize the description of gravity with quantum mechanics (see below). 

As mentioned above, phenomenologically, supersymmetry is needed to ensure the rendez-vous of coupling constants 
at about $10^{16} ~GeV$ \cite{Amaldi}. Supersymmetry does also provide many candidates for the 
{\it dark matter} (see Refs. \cite{Einasto,Roos} for recent reviews) of the Universe, 
given that it contains in its spectrum many neutral long lived weakly interacting massive particles (WIMPS); 
the linear combination of supersymmetric particles with the mentioned features is usually referred to as {\it neutralinos}.

\vskip 0.4cm

\centerline{{\bf Superstring/M-Theory Unification}}

\vskip 0.4cm

The basic assumption of string theory is that the fundamental building blocks of reality are not particles, but rather extended one 
dimensional objects, {\it quantum strings}. These quantum strings can be open or closed, and particles correspond to 
the modes of excitation of the strings. 
Open strings ask for ``branes" of any dimensionality not just {\it membranes}, a two dimensional object, to ``support" their ends. A striking feature of 
superstring theory is that at these contact points the 
interactions correspond to those of supersymmetry gauge theories. On its hand, closed strings admit the graviton 
in its spectrum, besides a scalar particle, the dilaton. 
Thus, a rather model independent prediction of string theory is that the emerging effective gravity 
theories arising from it are scalar-tensor theories of gravity 
with higher-order curvature terms \cite{Many}.      

The obstacles encountered in the development of the Kaluza-Klein approach seemed, as discussed above, insurmountable till 
the ``first string revolution" in 1984. Indeed, it was then shown by Green and Schwarz that higher order curvature terms do 
allow for non-trivial gauge fields configurations after compactification of $D=10$ down to $4$ dimensions, 
but also that to ensure the mutual cancellation of gauge and gravitational anomalies the only GUT gauge groups admissible are 
$E_8  \otimes E_8$ or $SO(32)$ \cite{GS84}. This breakthrough took place in the context of the supersymmetric string 
theory whose consistency (Lorentz symmetry and unitarity) requires  $D=10$ space-time dimensions 
(the bosonic string demands $D=26$ \cite{Polyakov}). 

Actually, in its very first avatar, string theory was proposed to describe hadronic physics. However, the persistent 
appearance of massless vector and tensor states in its spectrum and the fact that $D>4$ made the approach 
untenable for the description of hadrons. 
The first suggestion that string theory should instead be regarded as a unified theory of all interactions was put forward in 1974 by  
Scherk and Schwarz, based on the fact that the massless vector and tensor particles interact precisely as 
Yang-Mills gauge fields and the graviton \cite{SS74}. The former feature could be achieved by assuming 
that the fundamental length scale of the theory 
$L \simeq \alpha'^{1/2} = T^{-1/2}$, where $\alpha'$ is the so-called Regge slope and $T$ the string 
tension, should be identified with the Planck length, $L_P= \sqrt{G \hbar/c^3} = 1.6 \times 10^{-35}~m$. 

The ``first string revolution" did suggest a promising scenario to the understanding of our world. Starting from 
the $E_8  \otimes E_8$ $10$-dimensional superstring theory, the so-called heterotic string theory \cite{Gross}, 
it is natural to demand that the dimensional reduction process down to $4$ dimensions should preserve supersymmetry. This requirement 
turns out to be quite restrictive, as it 
demands that $6$ extra space dimensions are compact, have a complex structure, no Ricci curvature and an $SO(3)$ 
holonomy group. That is, this compact space must be a Calabi-Yau manifold 
\cite{CHSW}, a possibility that was thought to have opened the way to explain the origins of the SM. 

A ``second string revolution" has emerged from the discovery of the deep connection between all string 
theories. This is achieved through the so-called $S$ and $T$ dualities and the existence of an 
encompassing master theory, {\it M-theory}, which at low energies can be described by a $D=11$, $N=1$ supergravity theory \cite{Witten5}. 
The difficulty of M-theory is that it admits a huge number of solutions, about $10^{100}$ or greater --- and  
every possible value for the cosmological constant and coupling constants. 
The space of all such string theory vacua is often referred to as the {\it landscape} 
\cite{BoussoPolchinski00}.

In this context, a quite radical scenario emerges, namely that the 
multiple vacua of string theory is associated 
to a vast number of ``pocket universes'' in a single large {\it multiverse}. These pocket universes, 
like the expanding universe we observe around us, are all 
beyond any observational capability, as they lie beyond the cosmological horizon \cite{Susskind}. 
The implications of these ideas are somewhat disturbing: the vacuum that corresponds to our Universe 
must arise from a selection procedure, 
to be dealt with via anthropic or quantum cosmological considerations. That is to say that our existence somehow plays 
a role in the selection process. If, from one hand, the vast number of vacua in the landscape ensures 
the reality of our existence, a selection process must be evoked. One refers to the {\it anthropic landscape}, 
when the vacuum selection is based on anthropic considerations. 
This interpretation is not free from criticism: indeed,  it has been pointed out, for instance, 
that the impossibility of observing a multiverse implies that its scientific status is 
questionable --- It is in the realm of metaphysics, rather than of physics \cite{GEllis06}. This situation 
could be altered if the universes could interact. This possibility has been suggested in order to tackle the 
cosmological constant problem \cite{Bertolami2007a}.

\vspace{0.4cm}

\centerline{{\bf Unification in Cosmology}}

\vspace{0.4cm}
 
Cosmology is a particularly fruitful testing ground for unification ideas, given that the description of the Universe's history and 
evolution requires the integration of all physical knowledge. In the last few decades, the Hot Big Bang (HBB) model 
(see {\it e.g.} Ref. \cite{Bertolami2006d} for an extensive discussion, even though in Portuguese),
which is based most fundamentally 
in general relativity and quantum field theory, nuclear physics, statistical mechanics, {\it etc.}, has acquired the status of a paradigm. Indeed, the HHB model 
harmonizes all known observational facts, provided one admits the existence of states beyond the SM:  {\it dark energy}, which in 
its simplest form can be just a (fairly small) cosmological constant and {\it dark matter}. The HHB requires, at very early times, 
a period of accelerated expansion called {\it inflation} (see {\it e.g.} Ref. \cite{Olive} for an extensive discussion).
  
Inflation reconciles cosmology with causality, solving the horizon, the homogeneity and the rotation problems. Inflation also suggests 
an elegant mechanism for structure formation, based on the ubiquitous quantum fluctuations that all fields are subjected to, 
and in particular, the scalar field responsible for inflation, the {\it inflaton}. In the context of GUTs, inflation can also 
prevent that magnetic monopoles dominate the dynamics of the Universe and lead to its collapse just after the Big Bang. 
Inflation naturally connects cosmology with GUTs, supergravity, superstrings and, in general, with the physics at the very early Universe  
(see {\it e.g.} Ref. \cite{Bertolami1987}).     
 
The recent discovery of the late acceleration of the Universe has lead to the necessity of considering the existence of dark 
energy, and likewise inflation, a putative scalar field, dubbed {\it quintessence}, to drive the accelerated acceleration 
(see Ref. \cite{Copeland} for a review). It is interesting that some 
{\it quintessential inflationary models}  
have been suggested in order to unify inflation and dark energy \cite{Peebles}. The unification of dark energy and dark matter 
has also been proposed in the context of the Chaplygin equation of state and its generalization \cite{Bento2002}, with quite interesting 
phenomenological features.

 \section{The future of unification }
 
It is quite difficult, if not impossible, to predict the details of future developments in physics, 
and even harder to speculate about their inevitable turns and inflections. 
The unity of Nature and successes in describing the Universe along the lines of our discussion   
strongly suggest that the unification principle is a fruitful methodological tool. We have also seen that the lack of 
experimental evidence and phenomenological guidelines may render sterile any premature attempts of unification.

In this respect, the 
physics community is eagerly waiting for any new piece of information arising from the LHC, dark matter searches, SuperKamiokande-III, 
Planck Surveyor, GLAST/Fermi experiment, {\it etc.}, to direct its efforts. Naturally, in what concerns for instance the LHC, the main goals    
are the detection of the Higgs boson and of states related with supersymmetric extensions of the SM. Other objectives 
include the search for 
new forces of Nature, for states associated with the existence of extra 
dimensions and new states such as unparticles and ungravity, the latter associated to a putative infrared scale invariance of the 
SM \cite{Georgi}. 

On the theoretical front, it is evident that a deeper grasp of the strong coupling 
regime of field theories is required in order to address fundamental problems,  
from the existence of glue balls in QCD to the vacuum selection in string theory. 

Another issue that deserves particular attention is whether the ultimate description of space-time requires more general 
algebraic structures such as, for instance, {\it noncommutative geometry} \cite{Connes}. It is known that the position of strings on a 
D-brane satisfies the algebra of noncommutative geometry for a constant non-vanishing Kalb-Ramond field  \cite{SeibergWitten,Schomerus}, 
which suggests a connection between string theory and more complex algebraic structures, yet to be fully unraveled. 
An interesting related question 
concerns the possibility of extending noncommutative geometry to the phase space \cite{Bertolami2005,Bertolami2006c,Bastos2008a}. This 
seems to be a particularly relevant issue, given that phase space noncommutative geometry has many interesting features at the level of 
quantum cosmology \cite{Bastos2008b} . These include the possibility of obtaining quadratically integrable wave functions that 
solve the Wheeler-DeWitt equation for a Schwarzschild black hole and whose associated probability of reaching the singularity 
vanishes \cite{Bastos2009}. 
   	
In the past, a rich lore of new phenomena was found while focusing on the implications of known fundamental symmetries and also on 
aspects and conditions of
their violation. This might be particularly important, given that quantum field theory and general relativity rely fundamentally 
on Lorentz invariance - in its local version in the case of general relativity. In quantum field theory, the 
{\it breaking of Lorentz invariance}  
ensues the {\it breaking of CPT symmetry}. Given that the quantum field theories that describe the known interactions are local and 
unitary, one should not expect CPT violating effects. However, in the context of string field theory, solutions that violate 
Lorentz invariance and CPT symmetry have been found \cite{KosteleckySamuel} and a great deal of research has been done in order to 
study the full range of phenomenological implications of these solutions \cite{Kostelecky} .  
	
In what concerns general relativity, the most basic symmetry is the {\it Equivalence Principle}, which comprises the 
{\it Weak Equivalence Principle} 
(WEP), {\it Local Lorentz Invariance} (LLI) and {\it Local Position Invariance} (LPI). Despite the fact that these 
underlying invariances hold with great precision \cite{Bertolami2006b}, it is quite possible that the Equivalence Principle is violated, 
most particularly by the dark sector (see Refs. \cite{Bertolami2009,Bertolami2007b}). Clearly, any evidence about the breaking of 
the Equivalence Principle and any of its underlying assumptions may imply the crumbling of general relativity as we know it. 
Depending on the nature of the uncovered evidence, one can draw quite specific conclusions: for instance, 
the breaking of the WEP might indicate 
the existence of one or more new interactions of Nature at the particular range under scrutiny. The breaking of the LPI may point to a 
dependence of the coupling parameters on position and/or on time. These dependences might, for instance, be associated with the 
existence of extra dimensions (see {\it e.g.} Ref. \cite{Turyshev} for a discussion and for an observational strategy for detection).       
 
Finally, we mention the {\it Strong Equivalence Principle} (SEP), which states that gravitational self-energy couples to gravity likewise all 
other interactions. The validity of the SEP implies that spacetime geometry is uniquely determined by the metric, as 
prescribed by general relativity. Thus, further testing general relativity and its alternatives, such as for 
instance scalar-tensor theories of gravity and effective models arising from string theory, is a crucial 
line of future research --- one that might open fruitful theoretical and observational perspectives 
(see {\it e.g.} Refs. \cite{Odyssey,SAGAS} for discussions on space missions to test gravity).

\subsection*{Acknowledgments}

\vspace{0.3cm}

\noindent We would like to thank the colleagues of the organizing committee,  
Jo\~ao Teixeira Pinto, Lu\'is Saraiva, Pedro Gir\~ao, Pedro Gon\c calves Henriques, Waldir Oliva e Jo\~ao Cara\c ca, 
for the invitation to contribute in the homage to Aureliano Mira Fernandes, and Catarina Bastos and Jorge P\'aramos for 
the critical reading of the manuscript.


\end{document}